\documentclass[]{sigplanconf}
\usepackage{amsmath}
\usepackage{graphicx}

\usepackage{url}
\usepackage{color}
\usepackage{listings}
\begin{document}

\titlebanner{Fourth Conference on Partitioned Global Address Space Programming Model (PGAS10)}        
\preprintfooter{Development and performance analysis of a UPC Particle-in-Cell code}   

\title{Development and performance analysis of a UPC Particle-in-Cell code}

\authorinfo{Stefano Markidis}
           {University of Illinois at Urbana-Champaign, IL USA\\ Katholieke Universiteit Leuven, Belgium}
           {s.markidis@gmail.com}
\authorinfo{Giovanni Lapenta}
           {Katholieke Universiteit Leuven, Belgium}
           {giovanni.lapenta@kuleuven.be}

\maketitle

\begin{abstract}
The development and the implementation of a Particle-in-Cell code written in the Unified Parallel C (UPC) language for plasma simulations with application to astrophysics and fusion nuclear energy machines are presented. A simple one dimensional electrostatic Particle-in-Cell code has been developed first to investigate the implementation details in the UPC language, and second to study the UPC performance on parallel computers. The initial simulations of plasmas with the UPC Particle-in-Cell code and a study of parallel speed-up of the UPC code up to 128 cores are shown.
\end{abstract}


\terms
Particle-in-Cell, UPC performance


\section{Introduction}
The Unified Parallel C (UPC) \cite{UPC}, a parallel extension of the ANSI C, is one of the most promising Partitioned Global Address Spaces (PGAS) programming languages for scientific applications. The PGAS languages emerged as alternative to MPI~\cite{MPI}, OpenMP~\cite{OpenMP} and POSIX threads for parallel programming of parallel computers. Two main features make the PGAS languages perfect candidates for scientific computing on parallel machines: the Single Process Multiple Data (SPMD) execution model and the global address space. First, the SPMD execution model, such as in MPI~\cite{MPI}, guarantees performance and good scaling on parallel environments. Second the global address space enables the creation of shared data among processors making message passing among processors not necessary, and code implementation of scientific applications on parallel computers easier. 

The scientific application under study is the simulation of plasma by a computational method, called {\em Particle-in-Cell}, to simulate fusion energy machines and space physics plasmas~\cite{Markidis}. In the Particle-in-Cell method, a set of computational particles moves as real particles in a self consistent electric field calculated by solving the Poisson equation. Because of the use computational particles, the Particle-in-Cell method shares features with the other common simulation techniques, such as Molecular Dynamics, Monte Carlo, and Agent-based simulations~\cite{Allen}. For this reason implementation techniques for Particle-in-Cell codes are of large interest for other computational techniques also.

Studies of the UPC performance can be divided in two categories. A class of benchmarks focuses on study of the communication and data movement primitives to analyze communication latencies and bandwidth~\cite{Zhang}. The second category regards to the overall UPC performance of applications and mathematical kernels~\cite{El}, such as the UPC implementation of the NAS parallel benchmarks (NPB)~\cite{Nasa}. The present study belongs to this second category.

A first Particle-in-Cell code has been written in the UPC language. This skeleton Particle-in-Cell code (presented in Appendix A) mimics very closely the parallel decomposition, the synchronization pattern, the data layout of the production code written in C++ and MPI developed by the Authors~\cite{Markidis}. The code was written to study possible development and performance issues before developing a major production Particle-in-Cell code in the UPC language. 

The paper is organized as follows. Section 2 presents the governing equations, the numerical schemes and the Particle-in-Cell algorithm. Section 3 focuses on development and implementation issues: it analyzes the parallel work decomposition, the data layout, and presents how the reduction operations were completed. Sections 4, 5 and 6 conclude the paper showing the simulation results regarding to a plasma instability, the parallel performance using up to 128 cores, and analyzes the parallel speed-up. The skeleton version of the UPC Particle-in-Cell code, that has been used for the simulations of this paper, is included in Appendix A.

\section{The Particle-in-Cell Algorithm}
In the current UPC implementation of the Particle-in-cell method, three approximations in the plasma model have been assumed. First, a one dimensional geometry is used: plasma particles are constrained to move along a line. Although this geometry is unrealistic in many cases, it is acceptable when the plasma is characterized by a main direction of motion. For instance, beam interactions are inherently one dimensional and are well described in the one dimensional geometry. Second an electrostatic formulation of the plasma is used. This means that magnetic field is not present and particles can be only accelerated by electric field created by charge separation. Third, it is assumed that only electrons move while the motion of plasma protons is neglected: because protons are 1836 times bigger than electrons, they do not move considerably over the time period of a typical simulation and can be considered motionless.

In the Particle-in-Cell method, plasma electrons and protons are mimicked by the computational particles to determine the evolution of the plasma system \cite{Hockney}. The trajectories are computed by solving the equation of motion for each particle. In order to update the particle velocity, the force acting on the particle must be calculated. In the Particle-in-Cell method, a grid is introduced and an average force at each grid point is calculated, instead of calculating the force directly from other particles (such as in Molecular Dynamics simulations). At each grid point the charge density is found by counting the particles belonging to that cell. The values of the electric field defined on the grid are computed by solving the field equation starting from the charge density. The electric field acting on a particle is then calculated taking the electric field of the cell the particle belongs to. In this way, the introduction of a grid leads to a computational cost of $O(N_p + N_g\log N_g)$ (where $N_p$ is the number of particles, $N_g$ is the number of grid points and the fastest field solver requires $O(N_g\log N_g)$ operations), avoiding the $O(N_p^2)$ operations of calculating directly the electric field from other particles.

The Particle-in-Cell method consists of an initialization and a computational cycle, composed by update of particle positions and velocities, calculation of the field on the grid and calculation of interaction between grid and particles (interpolation). At the beginning of the Particle-in-Cell simulation, the particles positions and velocities are sampled according to the particle distribution of the specific problem. For instance, a beam can be represented by a uniform distribution in space with all the particles having the same drift velocity $V0$. After the initialization, the computational cycle is repeated many times.

\subsection{Particle mover}
At each computational cycle, the trajectories of the $N_p$ computational particles are determined by solving the equation of motion for each particle $p$:
\begin{equation}
\label{eom}
v_p = \frac{d x_p}{dt} \quad  \quad \frac{dv_p}{dt} = \frac{q}{m}E_p \quad  \quad p =1...N_p,
\end{equation}
where $x_p, v_p$ are the particle position and velocity, $q, m$ are the charge and mass of the particle, $E_p$ is the electric field acting on the particle, and $N_p$ are the number of particles. Many integrators can be used to solve numerically the particle equation of motion (Eq.\ref{eom}). The numerical technique used in the UPC Particle-in-Cell is the {\em leap-frog} method \cite{Hockney}, also known as {\em Verlet} algorithm in the context of Molecular Dynamics simulations \cite{Allen}.  In this scheme, the particle position is evaluated at integer time levels $n, n+1$, while the velocity at half integer time levels  $n-1/2, n+1/2$:
\begin{eqnarray}
\label{num_1}
x_p^{n+1} = x_p^{n}  + v_p^{n+1/2} \Delta t \\ v_p^{n+1/2} = v_p^{n-1/2}  +  \frac{q}{m} E_p^n\Delta t .
\end{eqnarray}

\subsection{Field solver}
The field equation in the electrostatic case (no magnetic field) reduces to the Poisson's equation: 
\begin{equation}
\label{field}
d^2 \Phi/dx^2 = - \rho,
\end{equation}
where $\Phi$ is the electrostatic potential, and $\rho$ the charge density. The electric field can be calculated from the potential as derivative of the potential $\Phi$:
\begin{equation}
 E = -d \Phi /dx
\end{equation}
The Poisson Equation \ref{field} is numerically differenced with a second order accurate central difference. If the space is discretized in $N_g$ cells with $\Delta x$ width, each grid point $g$ ($g =0,1,...(N_g-2),(N_g-1)$) arise an algebraic equation that forms a line of the matrix of the linear system:
\begin{eqnarray}
\label{num_2}
\Phi_{g+1}^n -  2\Phi_{g}^n + \Phi_{g-1}^n = - \rho_g^n (\Delta x)^2  .
\end{eqnarray}
The $N_g$ equations above form a linear system. A matrix-free iterative linear solver, the Conjugate Gradient (CG) \cite{Saad}, is used in the current implementation of the UPC code. This particular method was chosen because it is in use in the Author's Particle-in-Cell production code. Once the potential $\Phi^n$ is known at each grid cell, the electric field on the grid $E_g^n$ is computed by central difference in space:
\begin{equation}
\label{num_3}
E_g^n = - (\Phi_{g+1}^n - \Phi_{g-1}^n)/(2 \Delta x )
\end{equation}
\subsection{Interpolation}
The field Equation \ref{num_2} requires to calculate the charge density at each grid point. This is accomplished by counting the number of particles belonging to the cell, multiplying this number by the charge $Q$ and dividing by the grid spacing ($Q/\Delta x$). In the current implementation of the UPC Particle-in-Cell code, each particle contribution to the charge density is split in two parts to reduce the numerical noise~\cite{Hockney}. Each particle contributes to the charge density of two cells: the cell, the particle belongs to, with a weight ($w_1 = (x_p - x_g)/\Delta x$) that is proportional to the distance between the particle and cell center positions, and the neighbor cell that receives the rest of the charge contribution with weight $w_2 = 1 - w_1$. The same mechanism is used in the particle mover stage to calculate the electric field $E_p^n$ acting on the particle $p$. The electric field is taken as a weighted combination of the electric fields defined on the grid points that are closest to the particle.

\subsection{Algorithm summary}
In summary, the Particle-in-Cell algorithm is organized as follows. The simulation parameters (particle positions velocities, and fields) are initially set up and a computational cycle is repeated many times. At each computational cycle, the new particle positions are calculated by using the first part of Equation \ref{num_1}, and the charge density on the grid is calculated on the grid by counting the amount of charge per cell (interpolation particle to grid). Once the charge density is known, the electrostatic potential is calculated by solving the linear system of Equation \ref{num_2}, and the electric field is computed as the derivative of the electric field. Finally the new particle velocity is calculated by solving the second part of Equation \ref{num_1} as last step of the computational cycle.

\section{Development of the Particle-in-Cell in UPC}
A UPC Particle-in-Cell code has been implemented. The following sections present the development and implementation choices. A skeleton version of the Particle-in-Cell code is included in this paper in Appendix A to help the reader in understanding the Particle-in-Cell algorithm and the development issues.

\subsection{Parallel decomposition and synchronization}
Although the Particle-in-Cell method is a rather simple algorithm in the basic formulation, it poses at least two parallel designing challenges. 

First, because particles and grid points are coupled by the two interpolation steps (to calculate first the charge density on the grid from the particles, and then the electric field on the particle from the grid), particles and the grid points, that are close to the particles, need to be stored on the same process to avoid interprocess communication. This is achieved by dividing the simulation box in smaller domains, and placing the part of grid and particles belonging to that domain on the same process (Domain Decomposition). Once particles move from one domain defined on one process to the contiguous domain defined on a different process, the variables of these particles will be communicated to the different process. In the UPC Particle-in-Cell code each process has two communication buffers, that are filled with particles exiting the left and right sides. After the the particle positions have been updated, contiguous domains exchange the particle buffers. Moreover, the values of neighboring cells at the edge of the domain will be communicated to calculate the potential and electric field values. 

Second, because the Particle-in-Cell in its common formulation is a {\em time-driven} simulation, each stage of the Particles-in-Cell has to be synchronized over all the concurrent processes. Barriers need to be added in the code to ensure all the processes reach the same point before advancing to the next simulation step.

\subsection{Data layout}
Different techniques have been proposed to organize particle and field variables in memory to obtain efficient computer codes~\cite{Markidis2}. There are two main approaches: data can be organized in structure of arrays (SoA), where particles positions and velocities are stored in arrays, or in arrays of structures (AoS), where each particle position and velocity (and other particle data) is organized in a single particle structure. The first approach allow to use compiler auto-vectorization, while the second requires explicit vectorizing instructions coding. Moreover, the second approach leads to an increased cache performance. Because the main code developed by the Authors is based on the first approach, the UPC Particle-in-Cell code stores particles and field variables in arrays. This data layout results in enhanced performance of the code~\cite{Markidis2}. Memory in the PGAS languages is logically partitioned in shared and private memories, in a {\em two-level memory hierarchy} fashion. Shared memory is global and accessible by all the threads: the same memory address on each thread refers to the same memory location. This comes at the cost of an increased performance overhead due to hidden memory movement and communication (on distributed memory machine). Instead, the private memory is local in each thread: the same memory address on each thread refers to different memory locations. A {\em shared} type qualifier is used to declare a variable shared among threads, while all the variables are implicitly local in absence of it.  Moreover, the UPC language defines the concept of {\em affinity} as association between a shared variable and threads. Each element of data storage that contains shared objects has affinity to exactly one thread. The use of variables with affinity to the same thread improves the performance. It is important to consider affinity in order to avoid unnecessary data movement and communication (on distributed memory machines). When dealing with the Particle-in-Cell code, a concern is the memory layout of field quantities (charge density, electrostatic potential, and electric field) and particles quantities (positions and velocities). The simulation quantities are defined as {\em shared} variables in the current implementation. A one dimensional domain decomposition has been chosen to decide the {\em affinity} and divide the workload among the threads: contiguous computational cells with the information about the charge density, electrostatic potential and electric field are assigned to each thread. An array $np0$ keeps track of how many particles are located on each thread. Particle positions and velocities $xp0, vp0$ are stored as a one dimensional array with fixed length, and maximum block size. Two communication buffers $xout,vout$, one for particle exiting the left side and one for the right side, save the information of particle moving between contiguous processes. The field quantities $E$,$phi$ and $rho$ are shared among processes in the same way, and two additional cells $rho\_ghost$ provide temporary storage during the interpolation stage that will be communicated between neighbor domain.

\subsection{Reducing operations}
The iterative matrix-free CG solver of the UPC Particle-in-Cell method requires to calculate the inner product of shared arrays, whose blocks have affinity with different threads. The inner product operation is obtained by computing locally a contribution to the whole inner product, and reducing by adding the local values over all the processes. Different solutions were possible to perform parallel reduction in the CG solver. In this work the BUPC collective library \cite{UPC} based on the communication library (GASNet) is used. In particular the function {\em bupc\_allv\_reduce\_all((TYPE, TYPE value, upc\_op\_t reductionop))} is called at the beginning and at each iteration of the field solver.
\section{Verification Test}
\begin{figure*}
\centering
\includegraphics[height=5.6cm]{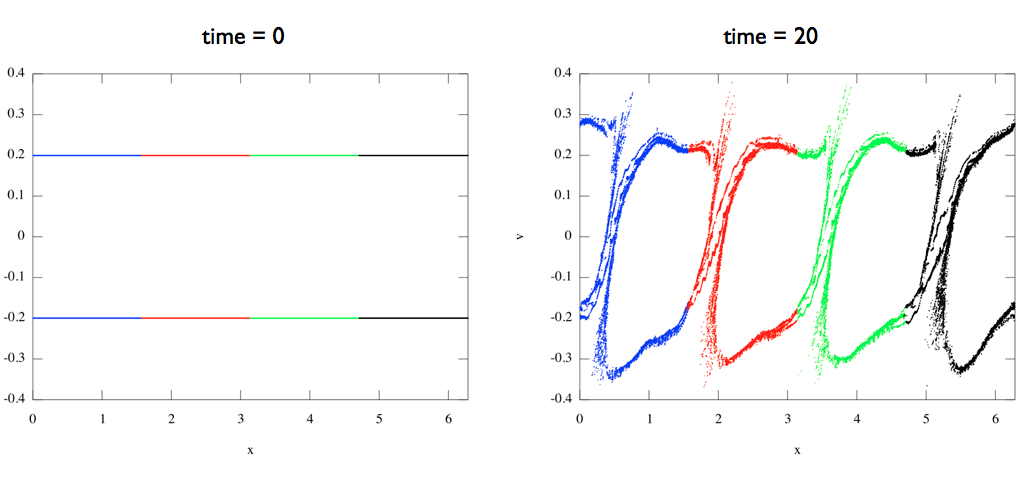}
\caption{Phase space (x,v) of the two stream instability with the UPC code. Two electron beams counterstream with drift velocities $+/- 0.2$ at $t=0$. The instability develops mixing the two beams in the phase space. The simulation has been completed using four cores. Each color represents the thread particles belong to.}
\label{twostream}
\end{figure*}

The code has been verified with the standard benchmark of the two stream instability \cite{Hockney}. The two stream instability is an important phenomenon occurring in space physics, in the injection systems for nuclear fusion machine, and in particle accelerators. In this problem, two electron beams flow initially in opposite directions in a neutralizing background of motionless protons. The two beams extinguish as result of the beam instability. Figure \ref{twostream} shows two time snapshots of the phase space ($x,v$ particles position-velocity space). In this plot, each particle represents a single point, whose the $x$ coordinate is its position and the $y$ coordinate is its velocity. The beams are moving at $V0=+/-.2/c$ (c is the speed of light in vacuum) and beam particles are uniformly distributed in space at time equal to zero. Therefore this distribution results in two lines at $V0=+/-.2/c$ in the phase space plot of Figure \ref{twostream}. The instability grows and mixes the two electrons beams, creating vortices in the phase space at time equal to 20 (time is normalized over the inverse of the plasma frequency). Beam particles, that were moving at velocity $V0=+/-.2/c$ initially, move with velocity close to zero or with opposite direction. The UPC simulation of Figure \ref{twostream} has been completed with four threads on four cores. The different particle colors in the phase space plot of Figure \ref{twostream} represent the particles variables affinity.

\section{Performance Analysis}

\subsection{Test environment}
The NERSC supercomputer Franklin\cite{Franklin} has been used for completing the performance tests. Franklin is a Cray XT4 massively parallel computer with 38,128 cores. Each compute node consists of A 2.3 GHz single socket quad-core AMD Opteron processor ({\em Budapest}) and 8 GB of memory (2GB per core). The code has been compiled with the Berkeley UPC compiler version 2.10.2 \cite{UPC} using the Cray's Portals low-level communications network API. The number of processes is provided at compiler time using the flag {\em -T=NUM} for optimization purpose. 

\subsection{Weak scalability tests}
The performance tests have been completed increasing the number of process cores and keeping the workload per core fixed (following the weak scaling notion). In all these simulations, 64 grid points, 64000 particles, and a simulation box 3.1415 long is assigned to each core. A Maxwellian plasma with thermal velocity 0.2 is initialized at beginning of the simulation, and 1000 computational cycles have been run. Because the single domain length is kept fixed, approximately the same amount of particles are communicated among processes in all the simulations. Moreover the iterative solver was forced to complete 150 iterations at each simulation. The number of iterations of the CG solver increases with the total number of grid cells at a given required convergence tolerance and it is important to ensure that each core runs the same number of iterations for performance test purpose. It is expected that increasing the number of cores, data movement and synchronization result in overhead and increased execution time.

The relative parallel speed-up has been taken as measure of the parallel performance. The smallest run was performed with only four cores (one compute node on {\em Franklin} supercomputer), and thus ideal scaling for the smallest run with four threads has been assumed.

\begin{figure}
\centering
\includegraphics[height=5.5cm]{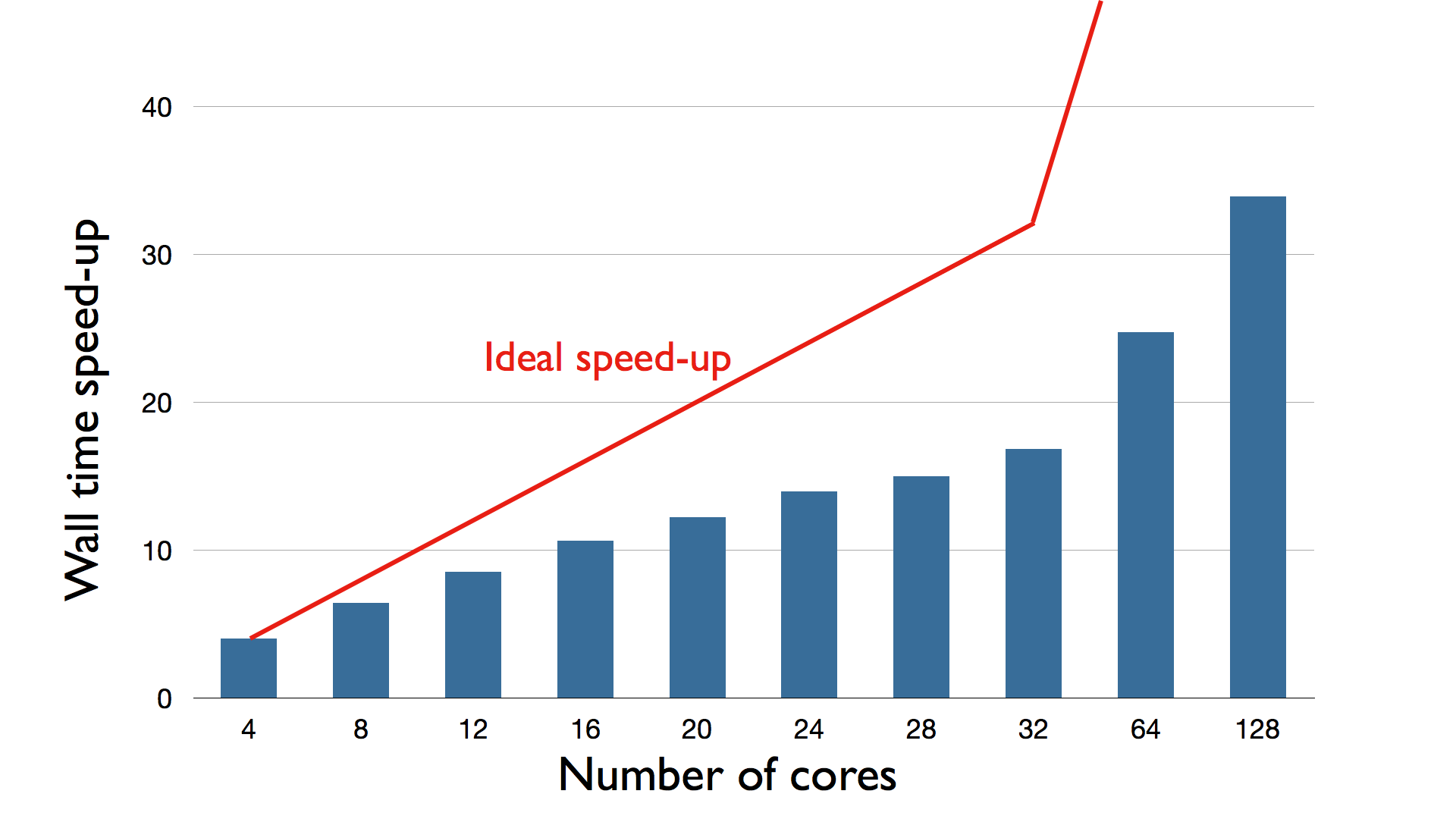}
\caption{Speed-up of the UPC Particle-in-Cell code. The red line shows the case of ideal scaling.}
\label{WallTimeSpeedUP}
\end{figure}

\begin{figure}
\centering
\includegraphics[height=5.5cm]{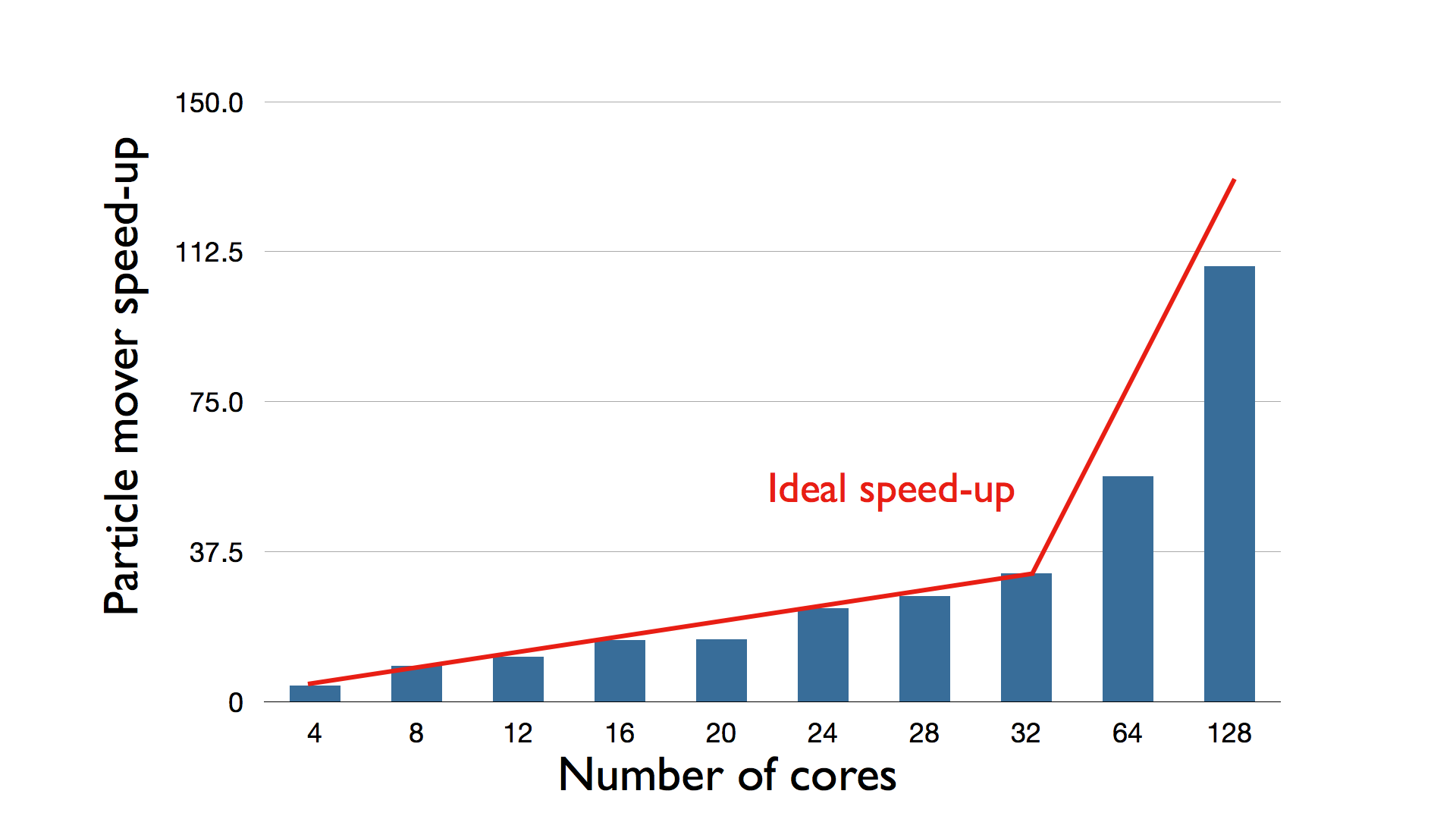}
\caption{Speed-up of the particle mover stage (Section 2.1). The speed-up and parallel performance are nearly optimal. A small degradation of the performance was expected as results of data movement (communication of particle data between neighbor processes and synchronization.}
\label{Particlemoverspeedup}
\end{figure}

\begin{figure}
\centering
\includegraphics[height=5.5cm]{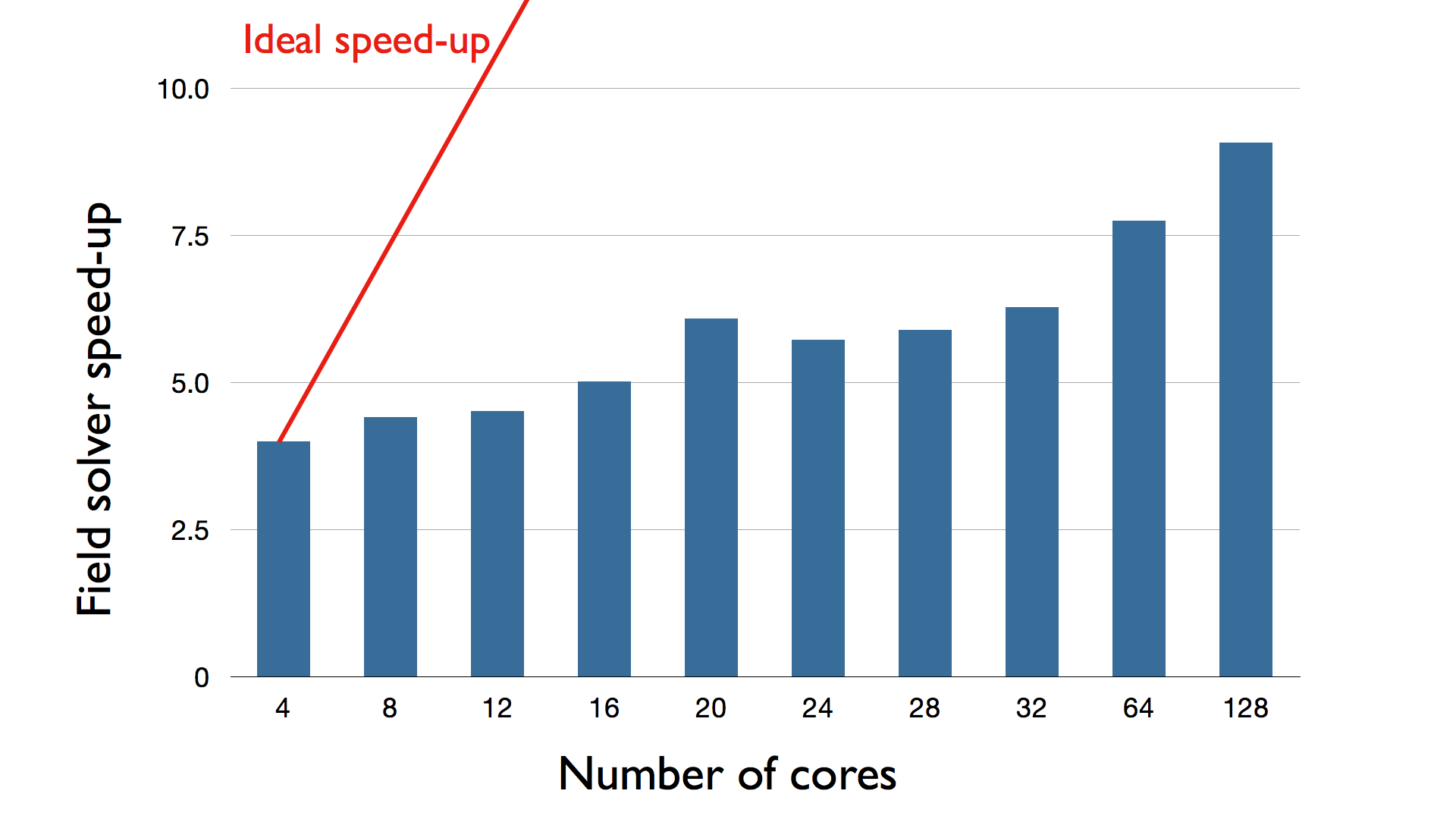}
\caption{Speed-up of the field solver stage (Section 2.2). In this case, almost no speed-up is observed increasing the number of cores. The performance degradation of the field solver is heavily reflected in the low speed-up of the UPC Particle-in-Cell method.}
\label{FieldSolverspeedup}
\end{figure}

\begin{figure}
\centering
\includegraphics[height=5.5cm]{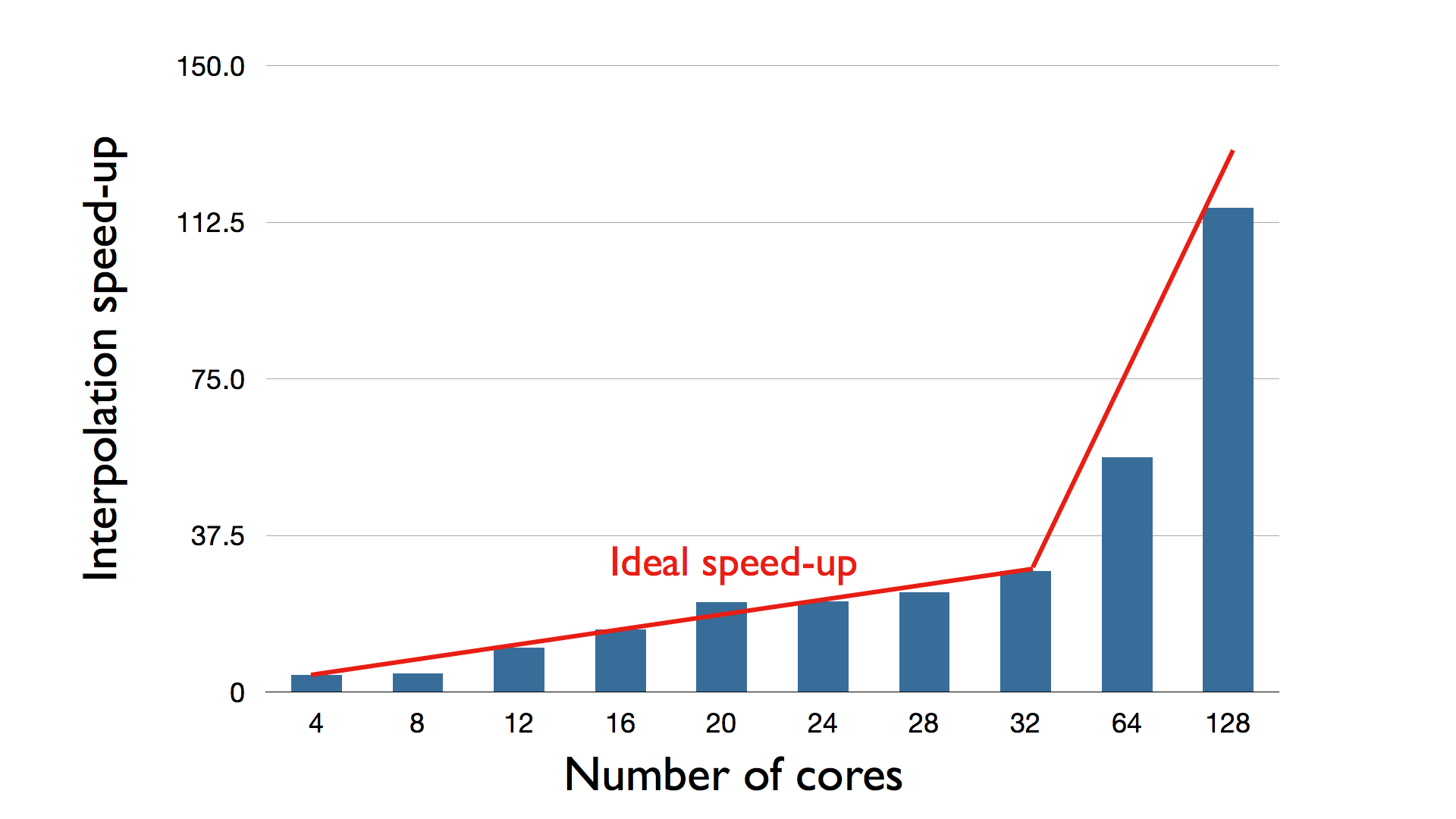}
\caption{Speed-up of the interpolation stage (Section 2.3). The speed-up is very close to the ideal scaling, as it is observed for the particle mover stage.}
\label{InterpolationSpeedUP}
\end{figure}

The speed-up of the UPC Particle-in-Cell code is presented first in Table \ref{UPCperformance} and then in Figure \ref{WallTimeSpeedUP}. The parallel speed-up does not increase considerably increasing the number of cores, and parallel efficiency degrades. A relative parallel efficiency of 26\% was recorded in the largest run with 128 cores.
\begin{table}
\caption{Parallel speed-up on the quad-core AMD Opteron processor (Budapest) from NERSC Franklin supercomputer with the Berkeley UPC compiler version 2.10.2 \cite{UPC}. The parallel speed-up increases moderately.}
\centering
\begin{tabular}{|c||c|c|}
\hline
\label{UPCperformance}
number of cores  & UPC Particle-in-Cell speed-up & ideal speed-up \\
\hline \hline
4  &  4 & 4\\ 
8  &  6.43 & 8\\ 
12 &  8.53 & 12\\ 
16 &  10.62 & 16\\ 
20 &   12.21 & 20\\ 
24 &   13.96 & 24\\ 
28 &   14.98 & 28\\ 
32 &    16.84 & 32\\ 
64 &   24.73 & 64\\ 
128 &  33.91 & 128\\ 
\hline
\end{tabular}
\end{table}
Figure \ref{WallTimeSpeedUP} shows that the parallel speed-up increases moderately with the number of cores on the $x$ axis. The red line represents the ideal scaling achievable in the best scenario where doubling the number of cores double the speed-up also. 
The speed-up of different parts of the Particle-in-Cell algorithm, that were presented in Section 2 (particle mover, field solver, and interpolation) have been analyzed individually to understand how different parts of the algorithm scale with the number of cores and where possible performance bottlenecks are.
Figure \ref{Particlemoverspeedup} shows that the parallel speed-up of the particle mover stage increases with the number of processes nearly ideally. Communication and synchronization results in a small performance decrease.
The cause of the performance degradation of the overall UPC Particle-in-Cell can be seen clearly in the Figure \ref{FieldSolverspeedup}, that shows the speed-up of the field solver stage. The speed-up is slightly increasing with the number of cores resulting in degraded parallel performance. The field solver performance bottleneck affects considerably the speed-up of the UPC Particle-in-Cell code.
The interpolation step performance results shows an excellent speed-up in Figure \ref{InterpolationSpeedUP} 
In summary, the UPC parallel performance decreases increasing the number of cores, with a speed-up equal to 34 for a run with 128 cores. The speed-up of the UPC Particle-in-Cell code is largely affected by the field solver speed-up that does not scale with the number of cores. On the opposite, both particle mover and interpolation step show nearly ideal scaling. 

\section{Conclusions}
A Particle-in-Cell code for plasma simulation has been developed and implemented in the UPC language. The numerical algorithm is first presented, and the implementation in UPC are discussed. The Particle-in-Cell code has been tested on the standard benchmark of the two-stream instability, and the performance of this initial implementation on parallel computer has been reported. Weak scalability tests showed a relative speed-up of 17, 25, 34 using respectively 32, 64, 128 cores. Parallel performance decreases considerably increasing the number of cores. 

An analysis of the parallel speed-up of different parts of the Particle-in-Cell code revealed that the performance degradation at high number of processes is only due to the field solver part of the algorithm. Two causes could have degraded the field solver performance. First, 64 grid points per process do not provide enough workload for one process, and the global communication (or global data movement) largely dominate the computation with no advantage from the parallel computation. Second, the UPC reducing operation strategy could probably be optimized. Because a large number of reducing operations were executed at each computational cycle, it is important reduction operations are completed in the fastest way for a given computer architecture. In fact, reducing operations to calculate the inner product are completed one time at the beginning of the field solver and then twice for each iteration. Thus, 301 reducing operations were completed at each Particle-in-Cell computational cycle in the test cases. On the contrary, the particle mover and the interpolations stages presented nearly optimal parallel speed-up. Note that these two stages are not {\em embarrassingly} parallel problems, and high parallel performance is not guaranteed {\em a priori}.

In conclusion, this study indicates the UPC language can be used efficiently for the development of Particle-in-Cell codes, especially for the particle mover and interpolation stages. However, the field solver performance bottleneck needs to be analyzed more deeply and removed before proceeding to simulations with very high number of cores.
\onecolumn
\appendix
\section{UPC Particle-in-Cell code}
The skeleton version of the UPC Particle-in-Cell code, that has been used for the performance tests, is presented here.
\begin{lstlisting}
#include <upc.h>
#include <bupc_collectivev.h>
#include <stdio.h>
#include <math.h>

#define NG_PER_PROC 64 // Number of grid points per process
#define NP_PER_PROC UPC_MAX_BLOCK_SIZE // Maximum number of particles per process
#define NP_PER_BUFFER 16384 // Maximum size of the communication buffer
#define NG NG_PER_PROC*THREADS // Total number of grid points
#define NP NP_PER_PROC*THREADS // Total number of particles

// Convert a double to an int (for interpolation)
#define dtoi(d) ((int)(d)) 
// Delete one electron from the particle array
#define DELETE_e()    np0[MYTHREAD]--; \
xp0[ip] =  xp0[start_particle + np0[MYTHREAD]]; \
vp0[ip] =  vp0[start_particle + np0[MYTHREAD]]; \
ip--
// Add one particle to the particle array
#define ADD_e()       xp0[start_particle+ np0[MYTHREAD]] = xp0[ip]; \
vp0[start_particle + np0[MYTHREAD]] = vp0[ip]; \
np0[MYTHREAD]++
// Add particle (exiting the left side) to the communication buffer
#define OUT_e_LEFT()  x_outLEFT[p_outLEFT[MYTHREAD]][MYTHREAD] = xp0[ip]; \
v_outLEFT[p_outLEFT[MYTHREAD]][MYTHREAD] = vp0[ip]; \
p_outLEFT[MYTHREAD]++
// Add particle (exiting the right side) to the communication buffer
#define OUT_e_RIGHT() x_outRIGHT[p_outRIGHT[MYTHREAD]][MYTHREAD] = xp0[ip]; \
v_outRIGHT[p_outRIGHT[MYTHREAD]][MYTHREAD] = vp0[ip]; \
p_outRIGHT[MYTHREAD]++

shared int np0[THREADS]; // number of particles in the process
shared[NP_PER_PROC] double xp0[NP]; // particle positions
shared[NP_PER_PROC] double vp0[NP]; // particle velocities

shared[NG_PER_PROC] double xg[NG]; // cell center positions
shared[NG_PER_PROC] double rho[NG]; // charge density
shared double rho_ghost_right[THREADS]; // rho ghost cell (right side)
shared double rho_ghost_left[THREADS];  // rho ghost cell (left side)
shared[NG_PER_PROC] double phi[NG]; // Potential
shared[NG_PER_PROC] double E[NG]; // Electric field

shared[NG_PER_PROC] double xkrylov[NG]; // CG solver solution array
shared[NG_PER_PROC] double r[NG]; // CG solver residual array
shared[NG_PER_PROC] double v[NG]; // CG solver helper array used in the CG solver
shared[NG_PER_PROC] double z[NG]; // CG solver helper array used in the CG solver

shared  int    p_outLEFT[THREADS]; // number of particles exiting the left side
shared  double x_outLEFT[NP_PER_BUFFER][THREADS]; // positions of particles exiting the left side
shared  double v_outLEFT[NP_PER_BUFFER][THREADS]; // velocities of particles exiting the left side
shared  int    p_outRIGHT[THREADS]; // number of particles exiting the left side
shared  double x_outRIGHT[NP_PER_BUFFER][THREADS]; // positions of particles exiting the left side
shared  double v_outRIGHT[NP_PER_BUFFER][THREADS]; // velocities of particles exiting the left side

int main() {
	// simulation parameter (each thread owns its own copy)
	int    nop_init = NG_PER_PROC*1000; // nop_init: number of particles per process
	if (nop_init > NP_PER_PROC){printf("** nop_init too large change NP_PER_PROC \n"); return(-1);}
	
	int nt = 1000;  // number of computational cycles
	double dt = 0.1; // time step
	double Lx = 3.1415*THREADS; double dx = Lx/((double)NG); // simulation box and grid spacing
	double tol = 1E-3; // solver tolerance
	int maxit = 150;   // maximum number of iterations for the CG solver
	double VT0 = 0.2; // particle thermal velocity
	double V0 = 0.0; // drift velocity
	double rho_init0 = 1.0; // initial density
	double qoms0 = -1; // particle charge to mass ratio
	
	double start_x = MYTHREAD*(Lx/((double)THREADS)); // first point of the thread domain
	double end_x   = (MYTHREAD+1)*(Lx/((double)THREADS)); //  last point of the thread domain
	int    start_cell = MYTHREAD*NG_PER_PROC;  // first cell index in the domain
	int	   end_cell   = (MYTHREAD+1)*NG_PER_PROC-1; // last cell index in the domain
	int    start_particle = MYTHREAD*NP_PER_PROC; // first particle index in the domain
	
	double Q0 = (rho_init0*Lx/((double)(nop_init*THREADS)))*(qoms0/fabs(qoms0)); // particle charge
	double w1, w2, Ex; // interpolation weights and Electric field
	double dx_part = (Lx /((double)(nop_init*THREADS+1)));
	double initial_error, dotZV, c, d, t; // solver variables
	int ic,ip,it,ig,ii,isolver,counter; // counters
		
	// initialize number of particles per thread
	np0[MYTHREAD] = nop_init;
	// initialize cell coordinates
	upc_forall(ic=0; ic<NG;ic++; &xg[ic]){xg[ic] = dx/2.0 + dx*ic;}
	// initialize initial particle position
	counter = start_particle;
	for (ig=start_cell;ig <= end_cell;ig++){
	    for (ii = 0; ii < (np0[MYTHREAD]/NG_PER_PROC); ii++){
		   xp0[counter] = ig*dx + (ii + .5)*(dx/(np0[MYTHREAD]/NG_PER_PROC));
		   counter++ ; } }
	// initialize initial particle velocity (Muller box algorithm to generate a Maxwellian)
	double r1, r2, r3, w, pp;
	for(ip=start_particle; ip < (start_particle + np0[MYTHREAD]); ip++){
		do {
			r1 = 2.0*rand()/((double)RAND_MAX) - 1.0; r2 = 2.0*rand()/((double)RAND_MAX) - 1.0;
			w = r1*r1 + r2*r2;
		} while ( w >= 1.0 );
		    w = sqrt((-2.0*log(w))/w);
		    r3 = rand(); if (r3/((double)RAND_MAX) < 0.5) vp0[ip] = V0 + VT0* r1 * w; 
		                 else vp0[ip] = -V0 + VT0* r1 * w; }
	upc_barrier;	
	// start the computational cycle
	for (it=0; it < nt; it++){
		// update particle positions
		for(ip=start_particle; ip < (start_particle + np0[MYTHREAD]); ip++) xp0[ip] += vp0[ip]*dt; 
		// clean the number of particles exiting left and right domain
	    upc_forall(ic=0; ic<THREADS;ic++; &p_outLEFT[ic]){p_outLEFT[ic] = 0; p_outRIGHT[ic] = 0;}
		// fill the communication buffers
		for(ip=start_particle; ip < (start_particle + np0[MYTHREAD]); ip++){
		   // apply the periodic boundary conditions
		   if (xp0[ip] < 0){
					xp0[ip] += Lx; OUT_e_LEFT(); DELETE_e();
		   } else if (xp0[ip] > Lx){
					xp0[ip] -= Lx; OUT_e_RIGHT(); DELETE_e();
		   } else {
			   if      (xp0[ip] > end_x  ){ OUT_e_RIGHT(); DELETE_e(); } 
			   else if (xp0[ip] < start_x){ OUT_e_LEFT() ; DELETE_e(); }
		   }
		
		} // end of the cycle
		upc_barrier;
		// get particles from the right side
		for (ip = 0; ip < p_outLEFT[(MYTHREAD+1)%THREADS]; ip++){
			xp0[start_particle + np0[MYTHREAD]] = x_outLEFT[ip][(MYTHREAD+1)%THREADS]; 
			vp0[start_particle + np0[MYTHREAD]] = v_outLEFT[ip][(MYTHREAD+1)%THREADS];
			np0[MYTHREAD]++; }
		upc_barrier;
		// get particles from the left side
		for (ip = 0; ip < p_outRIGHT[(MYTHREAD-1 + THREADS)%THREADS]; ip++){
		    xp0[start_particle + np0[MYTHREAD]]  = x_outRIGHT[ip][(MYTHREAD-1 + THREADS)%THREADS]; 
			vp0[start_particle + np0[MYTHREAD]]  = v_outRIGHT[ip][(MYTHREAD-1 + THREADS)%THREADS];
			np0[MYTHREAD]++; }
		upc_barrier;
		// set densities to zero 
		upc_forall(ic=0; ic<NG;ic++; &rho[ic]) rho[ic] = 0.0; 
		upc_forall(ic=0; ic<THREADS;ic++; &rho_ghost_left[ic]){
			 rho_ghost_left[ic] = 0.0; rho_ghost_right[ic] = 0.0;}
		// add background ions 
		upc_forall(ic=0; ic<NG;ic++; &rho[ic]) rho[ic] += rho_init0;
		// Interpolation particle to grid
		for(ip=start_particle; ip < (start_particle + np0[MYTHREAD]); ip++){
			ig = dtoi(xp0[ip]/dx+0.5);
		    if(ig == start_cell)        {w1 = (xg[start_cell]-xp0[ip])/dx; w2 = 1 - w1; 
										rho[start_cell]+=w2*Q0/dx;   rho_ghost_left[MYTHREAD]  +=w1*Q0/dx;}
			else if (ig == (end_cell+1)){w1 = (xp0[ip] - xg[end_cell])/dx; w2 = 1 - w1; 
										rho[end_cell]  +=w2*Q0/dx;   rho_ghost_right[MYTHREAD] +=w1*Q0/dx;}
			else                        {w1 = (xg[ig]-xp0[ip])/dx;         w2 = 1 - w1; 
				                        rho[ig]+=w2*Q0/dx;           rho[ig-1]+=w1*Q0/dx;}
		}
		upc_barrier;
		// communicate ghost cells
		rho[start_cell] += rho_ghost_right[(MYTHREAD-1 + THREADS)%THREADS];
		rho[end_cell]   += rho_ghost_left[(MYTHREAD+1)%THREADS];
		upc_barrier;
		// CG SOLVER
		upc_forall(ic=0; ic<NG;ic++; &xkrylov[ic]) xkrylov[ic] = 0.0;
		upc_forall(ic=1; ic<NG;ic++; &r[ic]){ r[ic] = -rho[ic]*dx*dx; v[ic] = r[ic];}
		if (MYTHREAD==0){r[0] = 0.0; v[0] = 0.0;} // BC on Phi
        pp =0.0; upc_forall(ic=0; ic<NG;ic++; &r[ic]) pp += r[ic]*r[ic];
		c = bupc_allv_reduce_all(double,pp,UPC_ADD);
		initial_error = sqrt(c);
        isolver = 0; 
		if (initial_error > 1E-16){
		  while(isolver < maxit){
			if (MYTHREAD==0)  z[0] = v[0]; // BC: PHI = 0 on BC
			if (MYTHREAD==(THREADS-1)) z[NG-1] = v[NG-2] -2*v[NG-1];
			upc_forall(ic=1; ic<NG-1;ic++; &z[ic]) z[ic] = v[ic-1] - 2*v[ic] + v[ic+1];
			pp = 0.0; upc_forall(ic=0; ic<NG;ic++; &z[ic]) pp += z[ic]*v[ic];
			dotZV = bupc_allv_reduce_all(double,pp,UPC_ADD);
			t = c/dotZV;
			upc_forall(ic=0; ic<NG;ic++; &xkrylov[ic]) xkrylov[ic] += t*v[ic];
			upc_forall(ic=0; ic<NG;ic++; &xkrylov[ic]) r[ic] -= t*z[ic];
            pp = 0.0; upc_forall(ic=0; ic<NG;ic++; &r[ic]) pp += r[ic]*r[ic];
			d = bupc_allv_reduce_all(double,pp,UPC_ADD);
			upc_forall(ic=0; ic<NG;ic++; &v[ic]) v[ic] = (d/c)*v[ic] + r[ic];
			c = d;
			isolver++;
		  }
		} else { } // converged at the first iteration
		upc_forall(ic=0; ic<NG;ic++; &v[ic]) phi[ic] = xkrylov[ic];
		upc_barrier;
		// calculate the electric field
	    upc_forall(ic=1; ic<NG-1;ic++; &E[ic]) E[ic] = - (phi[ic+1] - phi[ic-1])/(2*dx);
		if (MYTHREAD==0) E[0] = - (phi[1] - phi[NG-1])/(2*dx); // Periodic BC
		if (MYTHREAD==(THREADS-1)) E[NG-1] = - (phi[0] - phi[NG-2])/(2*dx); // Periodic BC
		upc_barrier;
		// 	calculated the new particle velocity
        for(ip=start_particle; ip < (start_particle + np0[MYTHREAD]); ip++){
		    ig = dtoi(xp0[ip]/dx + 0.5);
			if (ig==0)     {w1 = (xg[0]-xp0[ip])/dx;w2 = 1 - w1; Ex= w2*E[0] + w1*E[NG-1];}
			else if(ig==NG){w1 = (xp0[ip] - xg[NG-1])/dx; w2 = 1 - w1; Ex = w2*E[NG-1] + w1*E[0];}
		    else           {w1 = (xg[ig]-xp0[ip])/dx; w2 = 1 - w1; Ex = w2*E[ig] + w1*E[ig-1];} 
			vp0[ip] += qoms0*Ex*dt; 
		 }
	} // end of the simulation
	upc_barrier;
	return(0);
}
\end{lstlisting}
\twocolumn
\acks
This research used resources of the National Energy Research Scientific Computing Center, which is supported by the Office of Science of the U.S. Department of Energy under Contract No. DE-AC02-05CH11231.

The present work is supported by the Onderzoekfonds KU Leuven (Research Fund KU Leuven) and by the NASA MMS Mission Theory Support. The 
research leading to these results has received funding from the European CommissionÕs Seventh Framework Programme (FP7/2007-2013) under the grant agreement no. 218816 (SOTERIA project, www.soteria-space.eu).


\bibliographystyle{abbrvnat}


\end{document}